\documentclass[a4paper]{jpconf}
\usepackage{graphicx,color}
\usepackage[normalem]{ulem} 

\definecolor{red}{rgb}{0.8,0,0}
\definecolor{violet}{rgb}{0.4,0,0.4}
\definecolor{green}{rgb}{0,0.5,0.0}
\definecolor{navy}{rgb}{0.0,0.0,0.6}
\definecolor{orange}{rgb}{0.8,0.2,0.0}
\begin{document}
\title{Conventional and Unconventional Pairing 
and Condensates in Dilute Nuclear Matter
}

\author{John W Clark$^{1,2}$, Armen Sedrakian$^3$, Martin Stein$^3$,
Xu-Guang Huang$^4$, Victor A Khodel$^{1,5}$, Vasily R Shaginyan$^6$, 
Mikhail V Zverev$^{5,7}$}
\address{$^1$ Department of Physics and McDonnell Center for the
Space Sciences, Washington University, St. Louis, MO 63130 USA}
\address{$^2$ Center for Mathematical Sciences, University of Madeira,
Funchal, 9000-390 Portugal} 
\address{$^3$ Institute for Theoretical Physics, 
J. W. Goethe-University, D-60438 Frankfurt am Main, Germany}
\address{$^4$ Physics Department \& Center for Particle Physics 
and Field Theory, Fudan University, Shanghai 200433, China }
\address{$^5$National Research Centre Kurchatov Institute, Moscow, 123182, 
Russia}
\address{$^6$ Petersburg Nuclear Physics Institute, NRC Kurchatov Institute, 
Gatchina, 188300, Russia}
\address{$^7$ Moscow Institute of Physics and Technology, Dolgoprudny, 
Moscow District 141700, Russia} 

\ead{jwc@wuphys.wustl.edu}

\begin{abstract}
This contribution will survey recent progress toward an understanding 
of diverse pairing phenomena in dilute nuclear matter at small and moderate 
isospin asymmetry, with results of potential relevance to supernova 
envelopes and proto-neutron stars.  Application of {\it ab initio} 
many-body techniques has revealed a rich array of temperature-density 
phase diagrams, indexed by isospin asymmetry, which feature both 
conventional and unconventional superfluid phases.  At low density 
there exist a homogeneous translationally invariant BCS phase, a 
homogeneous LOFF phase violating translational invariance, and an 
inhomogeneous translationally invariant phase-separated BCS phase.  
The transition from the BCS to the BEC phases is characterized in 
terms of the evolution, from weak to strong coupling, of the pairing 
gap, condensate wave function, and quasiparticle occupation 
numbers and spectra.  Additionally, a schematic formal analysis 
of pairing in neutron matter at low to moderate densities is 
presented that establishes conditions for the emergence of both 
conventional and unconventional pairing solutions and encompasses 
the possibility of dineutron formation.
\end{abstract}

\section{Introduction}
This report serves to review and analyze a body of recent findings on 
the phase diagram of dilute nuclear matter, calculated over wide ranges 
of density, temperature, and isospin asymmetry.  Quantitative results will 
be presented for the temperature-density ($T-\rho$) phase diagram at 
baryon densities below about half the saturation density of isospin-symmetric 
nuclear matter, but pairing phenomena that may occur at somewhat higher 
densities will also be addressed.  The corresponding studies 
\cite{SC,RC,YE,II,BE}, both theoretical and numerical, focus attention 
on the emergence of unconventional as well as conventional pairing in 
the ${^3S}_1$-${^3D}_1$ (deuteron) channel as well as associated BCS-BEC 
crossovers.  Earlier work on these and closely related themes has been 
described in \cite{eagles,schmrink,alm,baldo,hstein,lombardo} and
more recently in \cite{mao,heckel,xghuang,jin,AS2013,fankro,III}.

Application of {\it ab initio} quantum many-body theory to this problem 
domain has the distinct advantage that, within the density regime 
considered in the numerical study, the two-body nucleon-nucleon (NN) 
interactions are well constrained by the NN phase-shift data and 
the properties of the deuteron.  Nor is the problem purely academic, 
as it is directly relevant to the matter existing in supernovae 
envelopes and proto-neutron stars (having relatively low temperatures 
and low isospin asymmetries) and in neutron star crusts (cold, with 
large isospin asymmetries).

The complex phenomenology of dilute nuclear matter, summarized in its
$T-\rho$ phase diagram determined over a range of isospin asymmetries, 
arises from three sources:
\begin{itemize}
\item[(1)] 
Pauli exclusion acting for fermionic species (nucleons, tritons, $^3$He, 
etc.) and Bose-Einstein condensation (BEC) of bosonic species (such as 
deuterons and alpha particles). These are most effective at high particle 
densities (but below nuclear saturation), low temperatures, and low mass
number of nuclear species.  
\item[(2)]
Dominance of the longer-range attractive component of the 
$NN$ interaction.
At low densities and not-so-low temperatures, this component is
responsible for the formation of tightly bound nuclear clusters 
(deuterons, dineutrons(?), tritons, alphas, ...), which can undergo 
BEC in the case of bosonic clusters.  At higher densities and low 
temperatures it is responsible for the formation of 
Cooper pairs with pairing gap $\Delta$.  At small isospin asymmetries 
the pairing is in the triplet ${^3S}_1$-${^3D}_1$ channel, 
whereas for large asymmetries the pairing is in the $^1S_0$ 
singlet channel.

\item[(3)] Isospin asymmetry, induced by weak interactions, producing 
a mismatch of neutron ($n$) and proton ($p$) Fermi momenta, giving 
rise to mixed superfluid/normal phases and unconventional pairing 
$-$ Cooper pairs with nonzero center-of-mass (CM) momentum (the 
so-called LOFF phase \cite{LOFF}) or deformed neutron and proton Fermi 
surfaces \cite{SCS}.
\end{itemize}

\section{Phase Diagram of Dilute Isospin-Symmetric Nuclear Matter}

We begin with a brief examination of the low-density phase diagram for
the fiducial case of isospin-symmetric nuclear matter with equal 
neutron and proton baryon densities, $\rho_n=\rho_p$, as explored 
in \cite{SC}.  A model of low-density nuclear matter based on a 
simplified two-nucleon interaction was considered which exhibits behavior 
generic to systems of fermions interacting via a short-range repulsion 
and a longer-range attraction.  Such behavior includes both (i) 
formation of clusters tightly bound in the medium at lower densities 
and higher temperatures, and (ii) Cooper pairing in a BCS state at 
lower temperatures and higher densities.  Specifically, a Malfliet-Tjon 
model with MF-III parametrization \cite{MT} was chosen for the $NN$ 
interaction, consisting of a central but spin-dependent superposition 
of inner repulsive and outer attractive Yukawas, fitted to $NN$ $S$-wave 
phase shifts and deuteron binding.  This interaction shows a strong 
pairing instability in the $^3S_1$ channel. 

To study the low-temperature superfluid phase, the BCS gap equation was 
solved self-consistently for the energy gap $\Delta(T)$ and the chemical 
potential $\mu$ below the critical temperature $T_{sc}$, with results
shown in Fig.~1.  Proceeding from higher to lower densities in the 
domain under study, conditions range from weak coupling (WC) to 
strong coupling (SC) as measured by the ratio $\Delta(0)/|\mu|$; 
a change of sign of the chemical potential from positive to negative 
is a signature of the WC $\to$ SC transition.  In the low-$\rho$ 
limit the gap equation reduces to the Schr\"odinger equation for 
the two-body bound state, with energy eigenvalue given by $2\mu$, 
which is naturally identified with the deuteron and subject to 
Bose-Einstein condensation.  Accordingly, this earlier study provides 
a model of the BCS-BEC transition from BCS Cooper pairing in the 
$^3{S_1}$ state to a Bose condensate of simplified deuterons.

\begin{figure}[tb]
\begin{center}
\includegraphics[width=11truecm,height=8.5truecm]{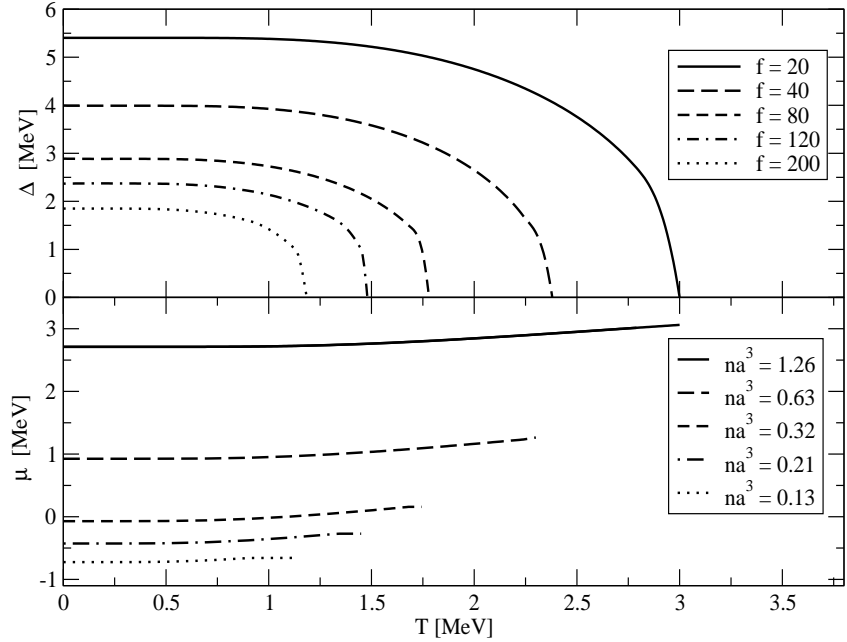}
\caption{Dependence of the pairing gap (upper panel) and
chemical potential (lower panel) on temperature for fixed values
of $f = \rho_0/\rho$, where $n \equiv \rho $ denotes the baryon density
and $\rho_0 = 0.16$ fm$^{-3}$ the saturation density of symmetrical 
nuclear matter.  Values of the dimensionless density parameter 
$n|a|^3$ assume a scattering length $a=5.4$ fm.}
\end{center}
\end{figure}

To extend the phase diagram to higher temperatures, the Lippmann-Schwinger 
and Faddeev equations were adapted to solve two- and three-nucleon 
bound-state problems in the presence of a dispersive fermionic background 
medium and attendant Pauli blocking effects.  Evolution of clustering 
into deuteron dimers and triton and helion trimers was followed under 
increasing temperature and/or decreasing density.  For small temperatures 
the quantum degeneracy is large and Pauli blocking strongly suppresses 
the binding energy of these clusters, which are quenched at a 
common critical temperature $T_{cc}$.

\begin{figure}[h]
\begin{center}
\includegraphics[width=11truecm,height=8.5truecm]{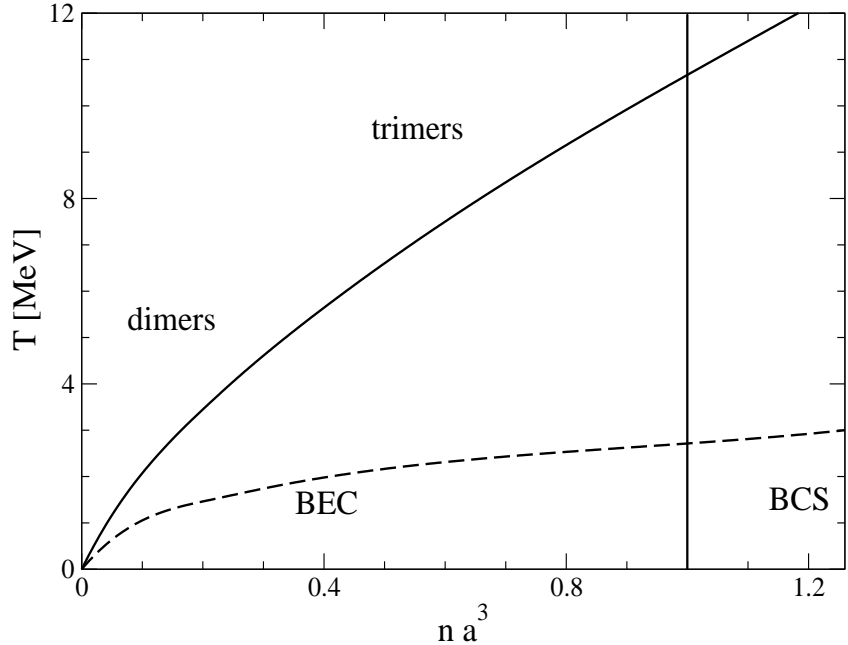}
\caption{
Phase diagram of dilute isospin-symmetric nuclear matter.
Solid line: Critical temperature for extinction of three-body
bound states; trimers and dimers exist above this line. 
Dashed line: Critical temperature for destruction of condensate.
Weakly coupled BCS superfluid exists below and far to right of 
vertical line, Bose-Einstein condensate (BEC) of tightly bound $np$ 
pairs exists below and far to the left.  Dimensionless density 
$n|a|^3$ (with $n \equiv \rho$) as defined in Fig.~1.
}
\end{center}
\end{figure}

The results on pairing and bound states are combined to produce
the schematic phase diagram in Fig.~2, showing several distinct 
regions in the $T-\rho $ plane:
\begin{itemize}
\item[G.]
The gaseous region above the solid critical line $T_{cc}(\rho)$ is 
populated by trimers, along with $np$ dimers at lower temperatures. 
\item[B.]
The low-temperature, low-density domain 
(lower left: 
$\rho|a|^3\ll 1$) 
contains a Bose condensate of tightly-bound deuterons.
\item[C.]
The low-temperature, high-density domain 
(lower right: 
$\rho|a|^3\gg 1$)
features a BCS condensate of weakly-bound Cooper pairs. 
\item[L.]
The domain between the two critical lines 
$T_{cc}(\rho)$ 
and 
$T_{sc}(\rho)$ 
contains normal nucleonic $\{p,n\}$ liquid.  
\end{itemize}

The superfluid phases labeled B and C are characterized by broken 
symmetry associated with the $\langle \psi\psi\rangle $ condensate.
The transition C to B does not involve symmetry changes $-$ it 
is a smooth crossover from the BCS to the BEC condensate.
B $\to$ L and C $\to$ L are second-order phase transitions related 
to the vanishing of the condensate along the line $T_{sc}(\rho)$.
The transition G $\to$ L (e.g., at vertical line) is characterized 
by an order parameter given by the fraction of trimers, which 
goes to zero at $T_{cc}(\rho)$ (tentatively second order).

\section{Effects of Isospin Asymmetry on the Phase Diagram}

The phase diagram of dilute nuclear matter becomes much more complex
upon introduction of isospin asymmetry as another control variable,
measured by 
\begin{equation}
\alpha = (\rho_n - \rho_p)/(\rho_n + \rho_p)
\end{equation}
in terms of neutron and proton number densities $\rho_n$ and $\rho_p$.
Unconventional superfluid and heterogeneous phases then emerge, largely 
dictated by mismatch of neutron and proton Fermi momenta $k_{Fn}$ and
$k_{Fp}$, which entails incomplete overlap of spherical neutron and 
proton Fermi spheres.  As is well known, this mismatch may be mitigated 
by deformation of the $n$ and $p$ Fermi spheres so as to increase 
their phase-space overlap \cite{SCS}.  Otherwise, phase-space 
overlap may be enhanced by the formation of Cooper pairs having non-zero 
CM momentum \cite{LOFF,SCS}.  This is the alternative explored quite 
thoroughly in the recent work that is the primary subject of this 
paper.  BCS pairing theory will be generalized to this case in the 
next section.  Increase of temperature is another option for compensating
the mismatch of $k_{Fn}$ and $k_{Fp}$, due to the smearing of both 
Fermi surfaces.  This effect, as well as the energetic advantage of 
dimerization at lower densities, gives rise to heterogeneous phases 
with superfluid and normal components.

The problem now has two energy scales: the  pairing gap $\Delta$ in 
the ${^3S}_1$-${^3D}_1$ channel, realistic $NN$ interactions now 
being employed, and the shift $\pm \delta \mu = \pm(\mu_n-\mu_p)/2$ 
between chemical potentials of neutrons and protons.
With increasing isospin asymmetry, $|\delta \mu|$ increases
from zero to values of order $\Delta$, and a sequence of
unconventional phases will appear.

\section{Gap Equation for Nonzero CM Momentum}

A BCS suitable gap equation allowing for Cooper pairs with nonzero
CM momentum may be derived in quasiparticle approximation in the 
framework of imaginary-time finite-temperature Green's functions 
and the Nambu-Gor'kov basis \cite{RC,II,YE}.  The resulting quasiparticle 
spectra are written in a general reference frame moving with CM 
momentum ${\bf Q}$ relative to a laboratory rest frame.

This {\it ab initio} many-body theory yields solutions of the form
\begin{equation}
G_{n/p}^{\pm} =
{ik_\nu \pm \epsilon_{p/n}^{\mp} \over
(ik_{\nu}-E_{\mp/\pm}^+)(ik_{\nu}+E_{\pm/\mp}^-)},
\end{equation}
\begin{equation}
F_{np}^{\pm} =
{{-i\Delta}\over{(ik_{\nu}-E^+_{\pm})(ik_{\nu}+E^-_{\mp})}},
\qquad
F_{pn}^{\pm} =
{{i\Delta}\over{(ik_{\nu}-E^+_{\mp})(ik_{\nu}+E^-_{\pm})}},
\end{equation}
for the normal and anomalous Green's functions, respectively,
where $k = (ik_\nu,{\bf k})$, with $k_\nu = (2\nu+1)\pi T$, 
$\nu$ being any integer. 
There are four quasiparticle spectral branches specified by
\begin{equation}
\epsilon^{\pm}_{{n\uparrow}/{\downarrow}} = E_S - \delta \mu
\pm E_A, \qquad
\epsilon^{\pm}_{{p\uparrow}/{\downarrow}} = E_S + \delta \mu
\pm E_A, \qquad
\end{equation}
where, with $a,r \in \{+,-\}$,
\begin{equation}
E_r^a  = \sqrt{E_S^2 + \Delta^2 } + r \delta\mu + a E_A,\qquad
E_S = (Q^2/4 + k^2)/2m^* - {\bar \mu} ,\qquad E_A =
{\bf k} \cdot {\bf Q}/2m^*.
\end{equation}
The isospin asymmetry enters through the parameter $\delta \mu =
(\mu_n - \mu_p)/2$, while ${\bar \mu}$ is the mean chemical potential.

In mean-field approximation, the anomalous self-energy (pairing gap) 
is expressed as
\begin{equation}
\Delta({\bf Q}) = {1 \over 4 \beta} \int {d^3 k' \over (2\pi)^3} 
\sum_\nu V({\bf k},{\bf k}')
{\rm Im}[F_{np}^+(k_\nu',{\bf k}',{\bf Q}) 
+ F_{np}^-(k_\nu',{\bf k}',{\bf Q})
- F_{pn}^+(k_\nu',{\bf k}',{\bf Q})
- F_{pn}^-(k_\nu',{\bf k}',{\bf Q})]
\label{anom}
\end{equation}
in terms of the above anomalous propagators, where $V({\bf k},{\bf k}')$
is the $np$ pairing interaction.  The next steps toward deriving the
gap equation as used in the numerical study of pairing in dilute
nuclear matter involve (i) evaluation of the Matsubara sum over $\nu$
in the above expression, (ii) performing a partial wave expansion, 
and (iii) restricting attention to the ${^3S}_1$-${^3D}_1$ channel, 
in which the appropriate $np$ interaction is denoted by 
$V_{l,l'}(k,k')$ with $l,l'=0,2$.  Thus one arrives at
\begin{equation}
\Delta_l(Q) = {1 \over 4}  \sum_{a,r,l'} \int {d^3 k' \over (2\pi)^3}
{{V_{l,l'}(k,k') \Delta_{l'}(k',Q)}
\over{2\sqrt{E_S^2(k') + \Delta^2(k',Q)}}} [1-2f(E_a^r)] ,
\end{equation}
wherein $f(E_a^r) = 1/[\exp(E_a^r/T) + 1]$ and $\Delta^2= (3/8\pi)\sum_l\Delta_l^2$.

Correspondingly, the partial number densities $\rho_{n/p}({\bf Q})$ of 
neutrons and protons are determined in terms of normal propagators by
\begin{equation}
\rho_{n/p}({\bf Q}) = {2\over \beta} \int {d^3 k \over (2\pi)^3}
\sum_\nu G_{n/p}^+(k_\nu,{\bf k},{\bf Q})
= 2 \int {d^3 k \over (2\pi)^3} {1 \over 2}
 \left[ ( 1 + \xi ) f(E_{\mp}^+) 
 + ( 1 - \xi ) f(-E_{\pm}^-) \right], 
\end{equation}
where $\xi = E_S/\sqrt{E_S^2 + \Delta^2}$. 

\section{Calculational Specifics and Interactions Assumed}

The coupled gap equations and the two density equations were
solved self-consistently for a bare pairing interaction 
in the ${^3S}_1$-${^3D}_1$ partial wave, as provided
by a phase-equivalent in-vacuum $NN$ interaction, namely 
by the Paris potential, thus implying Cooper pairing in 
the $S=1$, $T=0$ spin-isospin channel.  As needed, 
the nuclear mean field is modeled by a Skyrme density functional,
with SkIII and SLy4 parametrizations yielding nearly identical 
results.

Two simplifications are made:
\begin{itemize}
\item[(i)]
Polarization effects, i.e., 
medium modification of the input $NN$ interaction (due for example to 
virtual exchange of density and spin-density excitations) are neglected, 
although they are known to be important in some regions of the phase diagram.
\item[(ii)]
Apart from deuteron dimerization in the BEC phases, effects of nuclear 
clustering are not considered, although at somewhat higher temperatures
one expects substantial populations of tritons, $^3$He nuclei, 
and $\alpha$ particles, along with deuterons.
\end{itemize}
It should also be noted that ${^1}S_0$ Cooper pairing in the $S=0$, 
$T=1$ spin-isospin channel may mix and eventually replace 
${^3S}_1$-${^3D}_1$ pairing at asymptotically low $T$ (below 0.5 
MeV) and high asymmetry.  

\section{Free-Energy Minimization}

The phase at each point in the $T-\rho$ phase diagram is determined
by minimization of the free energy. 
In the case of pre-BEC homogeneous phases (perhaps translationally 
noninvariant), there are three possibilities:
(i) $Q = 0$, $\Delta \neq 0$ (BCS phase), (ii) $Q \neq 0$, 
$\Delta \neq 0$ (LOFF phase), and (iii) $Q = 0$, $\Delta =0$ 
(unpaired, normal phase).
The ground state is determined by minimization of the free energy
$F = E - TS$ of the superfluid (${\cal S}$) or unpaired normal phase 
(${\cal N}$) with respect to the parameter $Q$, where $E$ is the 
internal energy determined from the Hamiltonian and $S$ the entropy.  
Stability of the superfluid phase requires $F_{\cal S} < F_{\cal N}$.

As already indicated, there can also be a (pre-BEC) heterogeneous, 
phase-separated phase, denoted PS-BCS.  Its free 
energy takes the form of a linear combination of superfluid
and unpaired free energies,
$$
{\cal F}(x,\alpha) = (1-x)F_S(\alpha = 0) + xF_N(\alpha \neq 0),
$$
which is to be minimized with respect to the filling fraction $x$
of the unpaired component.  The net densities of $n/p$ per unit volume
are given by $\rho_{n/p} = (1-x)\rho_{n/p} ^{(S)} + x\rho_{n/p} ^{(N)}$.
In the pure $\cal S$ phase, $\rho_n^{(S)} = \rho_p^{(S)} = \rho^{(S)}/2$.  

\section{Overview of the Phase Diagram} 

In this section we present an overview of the diverse phases that
arise as the calculation proceeds from higher to lower densities
in dilute nuclear matter at chosen isospin asymmetries $\alpha \geq 0$.

\subsection{Conventional Phases}

As specified, the microscopic calculation yields a smooth crossover from 
the pure BCS phase to an asymptotic state corresponding to a mixture 
of a deuteron BEC and a normal gas of the left-over unpaired neutrons.  
The transition from BCS to BEC is identified by two criteria: 
(i) The average chemical potential $\bar \mu$ changes its sign from positive 
to negative values, and (ii) The coherence length $\xi$ of 
a Cooper pair becomes comparable to the interparticle distance $d$, 
$$
\xi \sim d = (3/4\pi \rho)^{1/3},
$$
as $\xi$ ranges from $\xi \gg d$ to $\xi \ll d$.

\subsection{Unconventional Phases: LOFF and PS-BCS} 

At $\alpha \neq 0$ the LOFF state can emerge due to the energetic 
advantage gained with a condensate that breaks translational symmetry: 
Cooper pairs that carry a nonzero CM momentum ${\bf Q}$ can compensate 
for the mismatch of neutron and proton chemical potentials.  The
calculations reveal the existence of a LOFF-state gap of isospin-asymmetric 
nuclear matter in a narrow regime at relatively low $T$ and 
relatively high $\rho$ values, having a maximum at finite $Q$,
implying maximum condensation energy for such pairs.  At large
$\alpha$, the maximum gap occurs at large values of $Q$. On the
other hand, at a given asymmetry, an increase of temperature shifts
the gap maximum and free-energy minimum toward smaller $Q$. 
With reduction of asymmetry, increase of temperature, and/or decrease 
of density, the BCS phase regains favor over the LOFF phase, with
PS-BCS phase having the advantage in the contest at lower temperatures.

This behavior relative to the LOFF phase is well understood in 
terms of the phase-space overlap of the Fermi surfaces of 
neutrons and protons, which increases with increasing 
temperature on the one hand, and with momentum $Q$ on the 
other. Similarly, the PS-BCS phase, in which a standard 
(${^3S}_1$-${^3D}_1$) BCS component coexists with a normal 
Fermi liquid made up of the excess unpaired neutrons, is
clearly favored energetically relative to pure BCS and LOFF 
phases at the lowest temperatures, where there is little 
benefit from eroded Fermi surfaces of from forfeit of 
translational invariance.

What remains is an account of how the BCS-BEC crossover is affected
by the existence of the unconventional nuclear LOFF and PS phases
at nonzero $\alpha$, under decrease of system density.  

\begin{figure}[tb]
\begin{center}
\includegraphics[width=12truecm,height=9truecm]{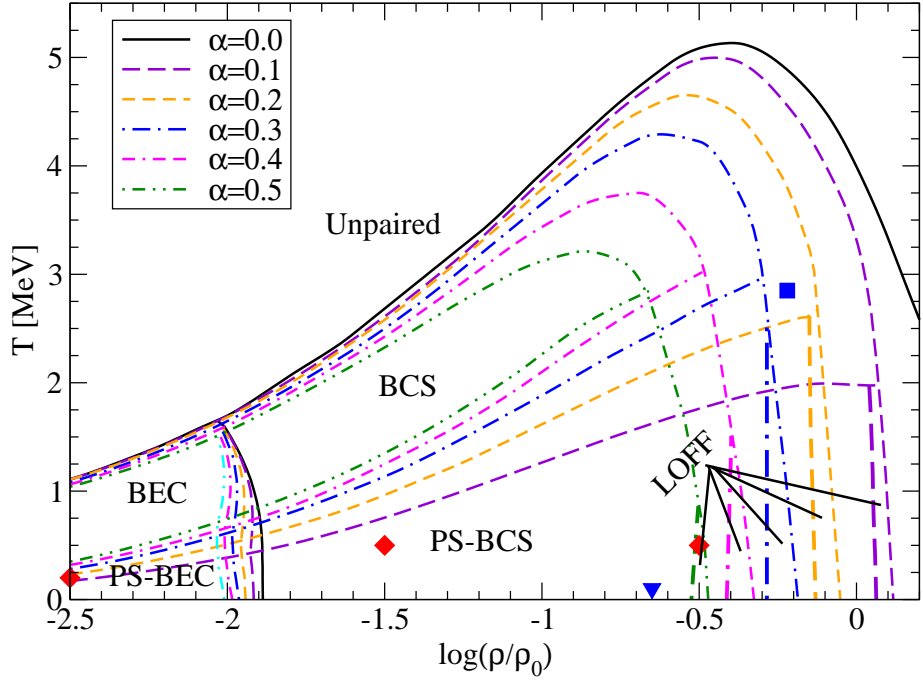}
\caption{
Phase diagram of dilute nuclear matter in the
  temperature-density plane at several isospin asymmetries $\alpha$ 
  as indicated for the plotted lines. Four phases are represented:
  the normal, unpaired phase, the BCS (BEC) phase, the LOFF phase, 
  and a PS-BCS (PS-BEC) phase. At each asymmetry there are two tri-critical 
  points, one always being Lifshitz point. For special values of 
  asymmetry these two tricritical points degenerate into single 
  tetra-critical point, exemplified for $\alpha_4= 0.255$
  (marked by blue square).  The LOFF phase disappears at the point 
  $\log(\rho/\rho_0) = −0.65$ and $T = 0$ for $\alpha  = 0.62$ 
  (marked by blue triangle).  Red triangles: representative points of 
  strong, moderate, and weak coupling, left to right.
}
\label{fig:phasendiagramm}
\end{center}
\end{figure}

\section{Inventory of Condensed (and Uncondensed) Phases}

With the density conveniently measured as $\log (\rho/\rho_0)$, 
results from the numerical calculations for different asymmetries 
$\alpha$ are plotted in the composite phase diagram of 
Fig.~\ref{fig:phasendiagramm}, which shows several distinct 
phases or domains:
\begin{itemize}
\item[I.]
We have renounced consideration of the formation of bound nuclear 
clusters with $A > 2$, which may occur at very low density and/or at 
high (but not excessively high) temperature.  The homogeneous unpaired 
(UP) phase is then always the ground state at high temperatures 
$T > T_{c0}$ in the restricted domain under study, where $T_{c0}$ 
is the critical temperature of the pairing transition at $\alpha = 0$ 
(conventional).
\item[II.]
The homogeneous isospin-asymmetric BCS phase is the ground 
state (denoted BCS) for all densities at intermediate temperatures
(conventional).
\item[III.]
The LOFF phase is the ground state in a narrow $T-\rho$ strip at
low temperatures and high densities (unconventional).
\item[IV.]
The domain of phase separation (PS), in either PS-BCS or PS-BEC 
realizations, appears at low temperatures (unconventional).
In the joint PS-BCS phase, one of the components is the 
isospin-symmetric BCS phase, while the other is the normal 
isospin-asymmetric phase.
\item[V.]
With decreasing density and intermediate or low temperatures, the phase
diagram shows two types of BCS to BEC crossovers from the asymmetrical
BCS phase to the BEC phase of deuterons and an embedded normal gas
of excess neutrons: (i) A transition between the homogeneous BCS/BEC 
phases at moderate temperatures (nominally conventional, but with a gas of 
leftover neutrons in the low-density limit) and (ii) a transition between 
the heterogeneous PS-BCS/PS-BEC phases at low temperatures (unconventional).  
By convention, boundaries between BCS (or PS-BCS) and BEC (or PS-BEC) 
phases are decided by the change of sign of the mean chemical potential 
${\bar \mu}$. In the phase diagram these appear as nearly vertical 
lines, insensitive to $\alpha$, seen in the low-$\rho$-low-$T$ corner 
of the phase diagram.
\end{itemize}

\section{Special Features of the Phase Diagram}

In terms of symmetries, four phases are identified: unpaired (UP), BCS/BEC, 
LOFF, and PS-BCS (PS-BEC).  The results of the calculations performed 
are consistent with the transitions between BCS to BEC or between 
PS-BCS and PS-BEC being smooth crossovers without change of symmetry.  
The superfluid/unpaired phase transition and the transitions between 
superfluid phases are second order, except for those between PS-BCS 
and LOFF phases, which are of first order.  At each nonzero isospin 
asymmetry $\alpha<0.62$, the phase diagram exhibits two tricritical 
points where a simpler pairwise coexistence terminates and three different 
phases coexist, e.g., BCS, PS-BCS, and LOFF. One of these is always a 
Lifshitz point.  For special values of asymmetry, the two tricritical 
points degenerate into a single tetracritical point; an example is shown
as the blue square in Fig.~\ref{fig:phasendiagramm}, occurring   
at $\log (\rho/\rho_0) = -0.22$ and $T=2.85$ for $\alpha = 0.255$.

\section{Beyond the Phase Diagram}

It is useful to distinguish three dynamical regimes:
\begin{itemize}
\item[$\bullet$]
The weak-coupling regime (WCR) corresponds to the high-density
limit where well-defined Cooper pairs are present.
\item[$\bullet$]
The strong-coupling regime (SCR) corresponds to the low-density
limit where well-defined deuterons are formed.
\item[$\bullet$]
In between: intermediate coupling regime (ICR).
\end{itemize}

The following properties of pairing and condensates in low-density 
nuclear matter have been examined in detail within the same 
calculational 
framework:
\begin{itemize}
\item[$\bullet$]
Temperature and asymmetry dependence of the pairing gap, with
comparison of BCS and LOFF phases.
\item[$\bullet$]
The kernel of the gap equation as the momentum space-wave function of 
the Cooper pair, with comparison of BCS and LOFF phases.
\item[$\bullet$]
Evolution of the Cooper-pair wave function from the WCR through the 
ICR to the SCR, i.e., evolution from BCS pairing to the BEC condensate 
of deuterons.
\item[$\bullet$]
Occupation numbers of neutrons and protons; their behaviors
from the BCS phase in WCR, through ICR, and on to BEC in SCR; 
LOFF in WCR.
\item[$\bullet$]
Quasiparticle excitations: dispersion relations for quasiparticle
spectra in the ${^3S}_1$-${^3D}_1$ BCS condensate and in the LOFF 
phase; evolution of spectral branches from WCR through ICR to SCR.
\end{itemize}

We next turn to selected samples of results from these informative
studies.  In figures providing results in the WC, IC, and SC
regimes, typical values have been selected at $(T,\rho)$ pairs
specified as follows.  WCR: $T = 0.5$ MeV, $\log(\rho/\rho_0)= -0.5$.
ICR: $T = 0.5$ MeV, $\log(\rho/\rho_0) = -1.5$.  SCR: $T= 0.2$ MeV,
$\log(\rho/\rho_0) = -2.5$.  

\subsection{Behavior of the Pairing Gap}

Results have been obtained for the pairing gap at density 
$\rho = 0.1$ fm$^{-1}$ (a) as a function of temperature for different 
asymmetry values, and (b) as a function of asymmetry for different 
temperatures (see Figs.~4 and 5).  When the possibility of a LOFF phase is taken
into account, these results, when plotted for each value of $\alpha$, 
reveal different regimes at relatively low and relatively high 
temperature.  The high-temperature segment corresponds to the 
BCS state, with standard temperature dependence of the gap.
By contrast, in the low-temperature region below the branch point, 
there are two competing phases: BCS and LOFF, with very different
temperature dependences of the gap function.

\begin{figure}[t]
\begin{minipage}{17.7pc}
\includegraphics[width=17.7pc,height=14pc]{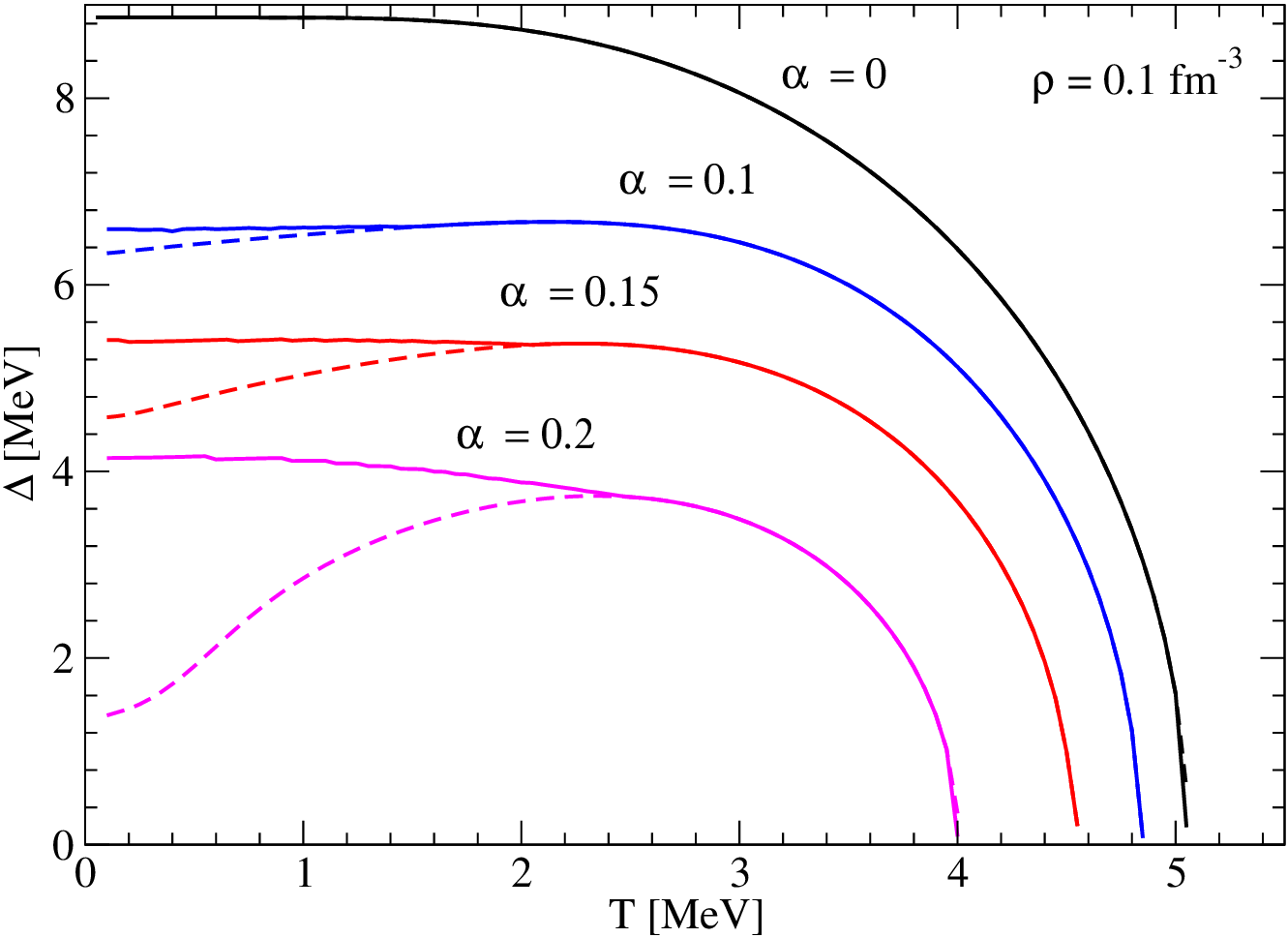}
\caption{
Gap as a function of temperature at constant density 
for asymmetry values $\alpha=0.0$
(black), $\alpha=0.1$ (blue), $\alpha=0.15$
(red), $\alpha=0.2$ (magenta).  Results allowing 
for the LOFF phase are traced by solid lines, those restricted to
the BCS phase, by dashed lines.}
\end{minipage}\hspace{3.5pc}%
\begin{minipage}{17.7pc}
\includegraphics[width=17.7pc,height=14pc]{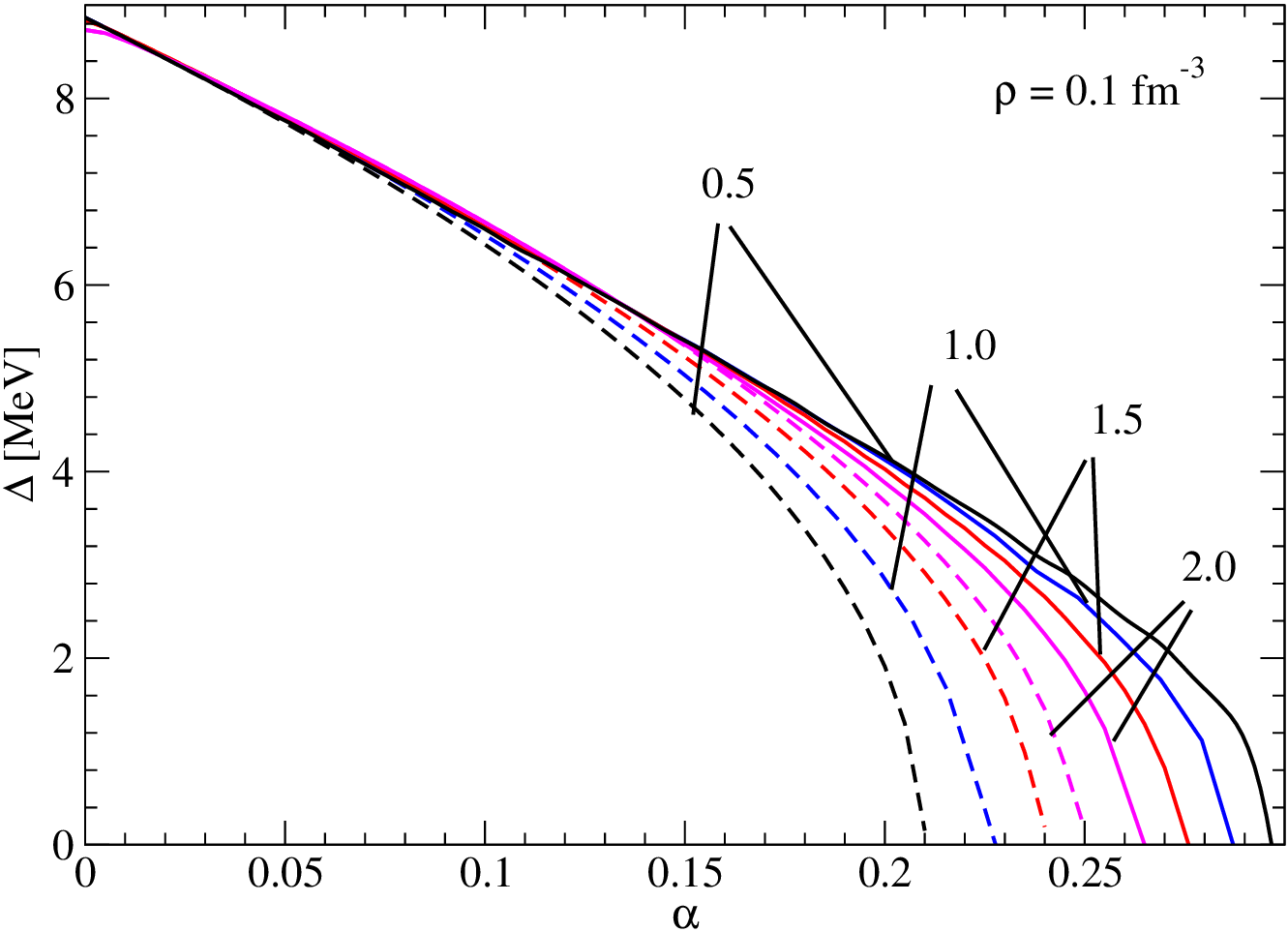}
\caption{
Gap as a function of asymmetry at constant density 
for temperature values $T=0.5$ (black),
$T=1.0$ MeV (blue), $T=1.5$ MeV (red), $T=2.0$ MeV
(magenta).  Results allowing for the LOFF phase are traced by
solid lines, those restricted to
the BCS phase, by dashed lines.}
\end{minipage} 
\end{figure}

The quenching of the BCS gap upon decrease of $T$ is caused 
by the loss of coherence among the quasiparticles as the 
thermal smearing of the Fermi surfaces is terminated, with 
the (unorthodox) consequence that for large enough asymmetries 
there exists a lower critical temperature $T_{c\downarrow}$.

In the plots of Fig.~5 showing $\Delta$ versus $\alpha$ for several
chosen temperatures and fixed density, there are two curves
for each $T$ value: one in the low-$\alpha$ regime where only
the BCS phase exists, and the other in the large-$\alpha$
regime where both BCS and LOFF are possible.  The LOFF
solution wins the competition in the latter region, since 
it provides larger gap values.

\begin{figure}[h]
\begin{minipage}{17.7pc}
\includegraphics[width=17.7pc,height=14pc]{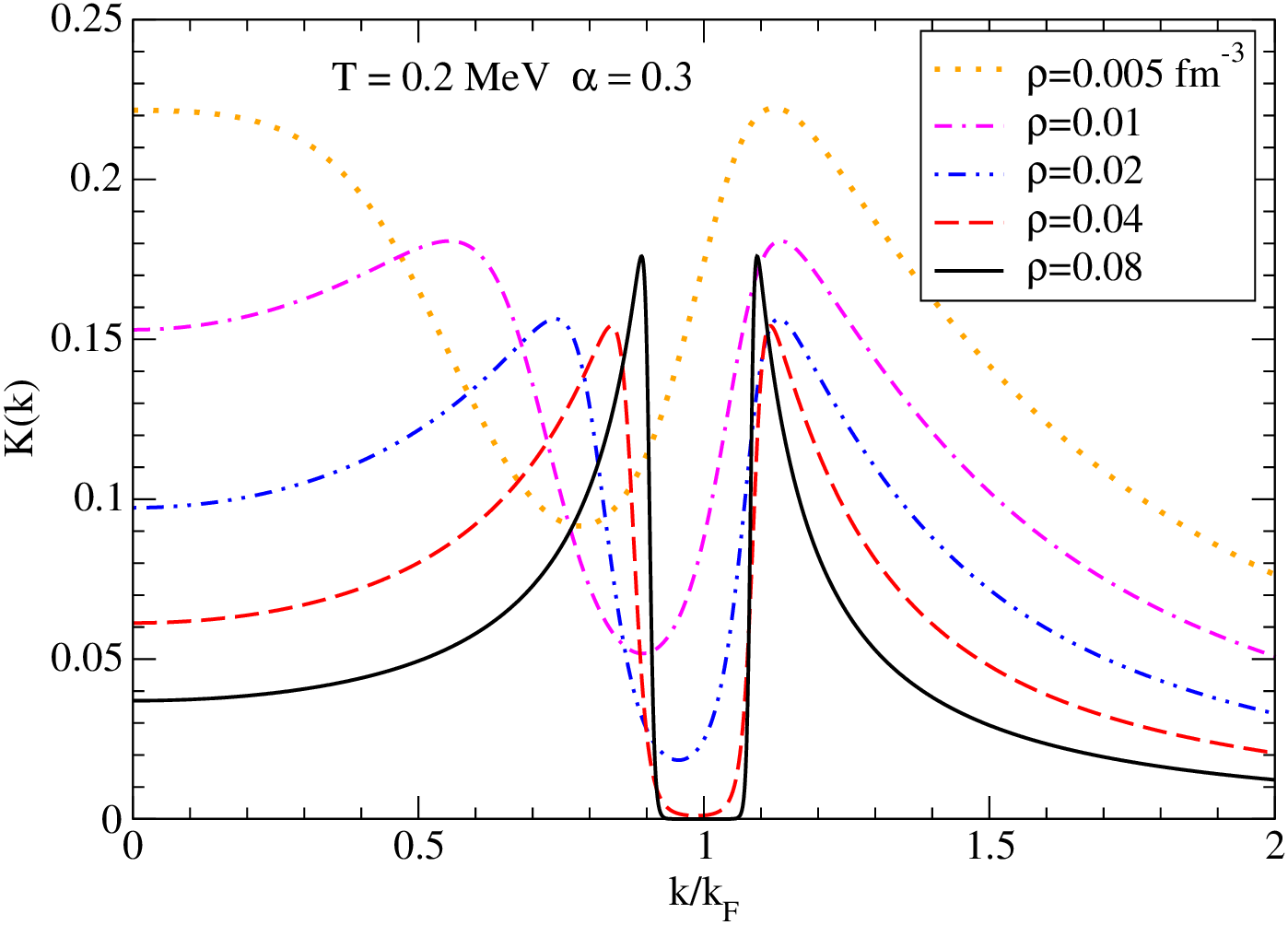}
\caption{Dependence of the kernel $K(k)$ on momentum in units
of Fermi momentum for fixed $T=0.2$ MeV, $\alpha = 0.3$, and
various densities indicated in the plot.  }
\end{minipage}\hspace{2pc}%
\begin{minipage}{17.7pc}
\includegraphics[width=17.7pc,height=14pc]{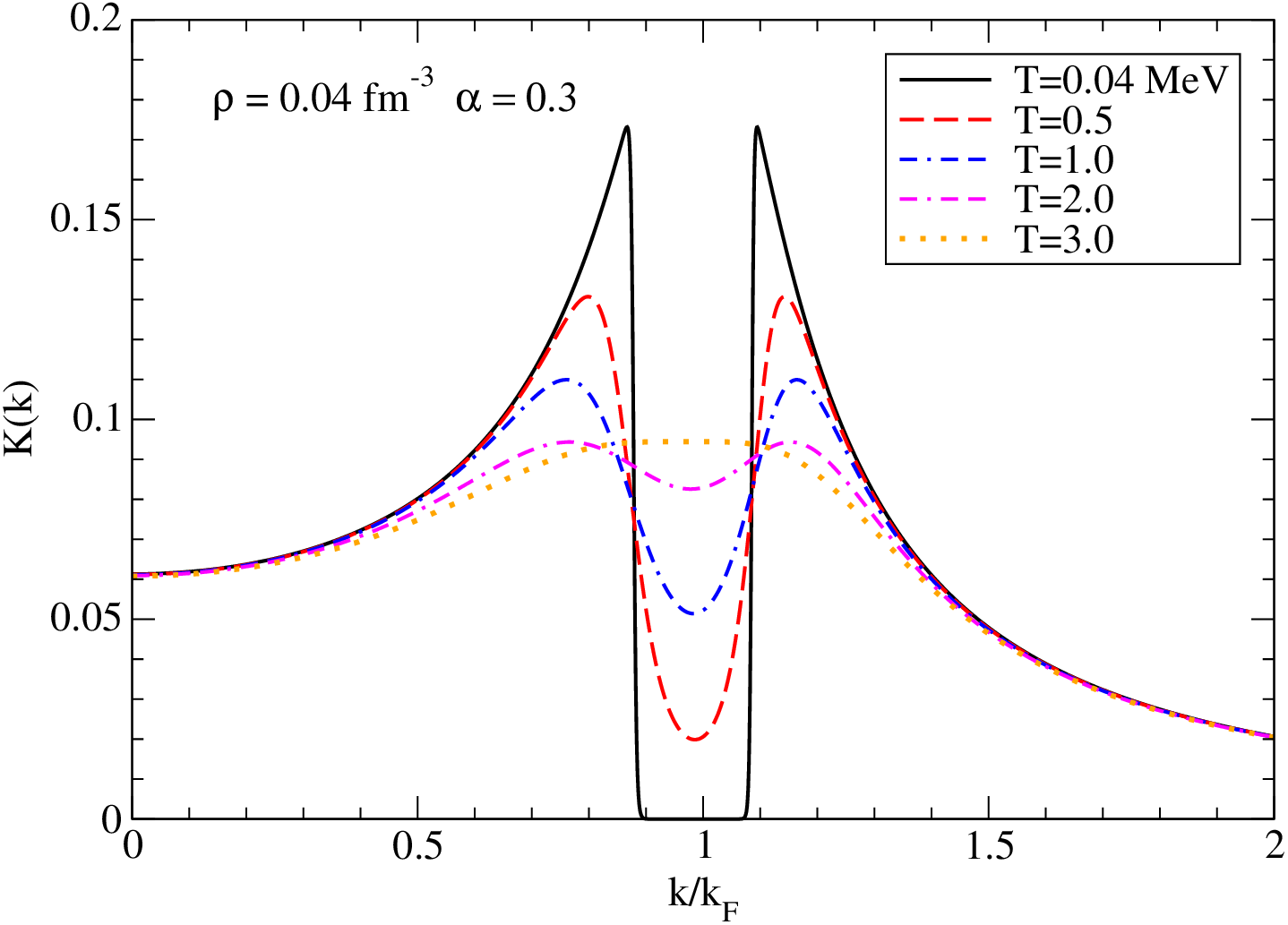}
\caption{Dependence of the kernel $K(k)$ on momentum in
units of Fermi momentum for fixed $\rho=0.04$ fm$^{-3}$,  
$\alpha=0.3$, and various temperatures indicated in the plot.  }
\end{minipage} 
\end{figure}

\subsection{Kernel of the Gap Equation and Cooper-Pair Wave Function}

The kernel of the gap equation,
\begin{equation}
K(k) =  {1 \over 2} \sum_{a,r}
\frac{1}{2\sqrt{E_{S}^2(k)+\Delta^2(k,Q)}}[1-2f(E^r_a)],  
\end{equation}
is the product of the imaginary part of the retarded anomalous 
propagator and the Pauli operator $P_r^a = 1-2f(E^r_a)$.  Physically, 
$K(k)$ can be interpreted as the momentum-space wave function 
of the Cooper pairs, since it obeys a Schr\"odinger-type eigenvalue 
equation in the limit of extremely strong coupling.  The Pauli 
operator is a smooth function of momentum with a maximum at the 
Fermi surface, where $E_S$ vanishes. In practice, contributions 
from the two relevant excitation branches with $r\neq a$ are kept; 
those from the cases $r=a$ are negligible.  Plots showing the
momentum dependence of the kernel for asymmetry $\alpha = 0.3$
and relevant values of density and temperature are provided
in Figs.~6 and 7.

\subsection{Cooper-Pair Wave Function and Correlation Length}

The superfluid coherence length $\xi$ is directly related to the
root-mean-square radius of the ${\bf r}$-space Cooper-pair wave
function, given by
\begin{equation}
\Psi({\bf r})
= {\cal N} \int {{d^3p}\over{(2\pi)^3}}
[K({\bf p},\Delta)-K({\bf p},0)]e^{i{\bf p} \cdot {\bf r}},
\end{equation}
where the factor ${\cal N}$ ensures unit norm for $\Psi({\bf r})$.
The mean-square radius of the Cooper pair is then defined as
\begin{equation}
\langle r^2 \rangle = \int\, d^3r r^2 |\Psi({\bf r})|^2,
\end{equation}
and its spatial extent, the coherence length, as
$\xi_{\rm rms} = \sqrt{\langle r^2 \rangle}$.  In traversing
the BCS-BEC crossover, the change in the coherence length tracks
the change of the condensate wave function.  The regimes of weak
and strong coupling can be identified by comparing the coherence
length, given in the BCS case by the familiar result
$\xi_a = \hbar^2 k_F/\pi m^*\Delta$, to the mean distance
$d = (3/4\pi \rho)^{1/3}$.  Detailed computation and analysis based
on these relations firmly establishes that in the BCS limit (WCR)
one is dealing with a coherent state whose wave function
oscillates over many periods characterized by $k_F^{-1}$.
In the opposite limit (SCR), the wave function is concentrated
around the origin, indicating that one is dealing with a
Bose condensate of strongly bound states: deuterons.
Plots illustrating the behavior of $r^2 |\Psi(\bf r)|^2$ versus $r$
in the different coupling regimes and for selected asymmetries
$\alpha$ are provided in Figs.~8 and 9.

\begin{figure}[t]
\begin{minipage}{18pc}
\includegraphics[width=17.7pc,height=16.6pc]{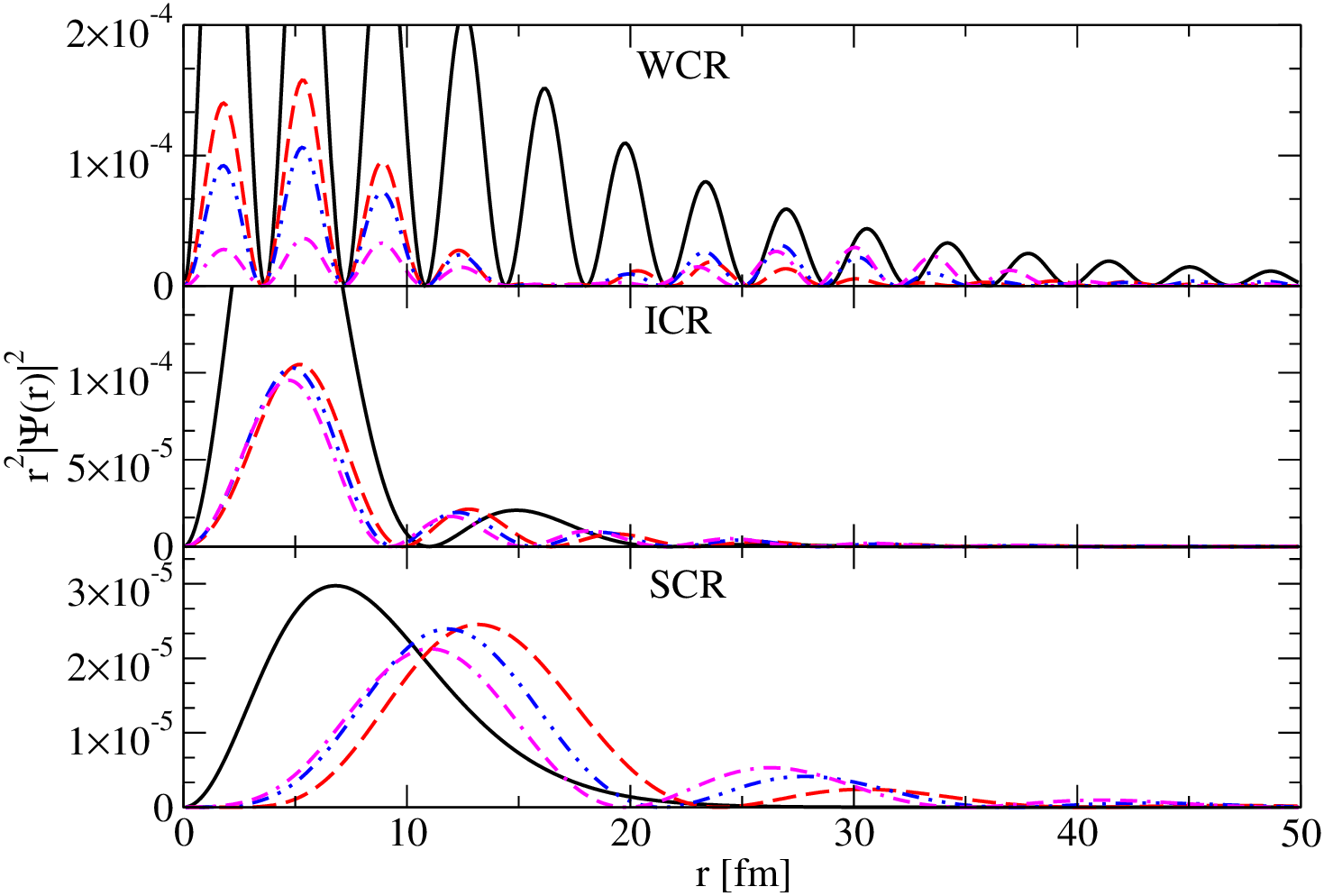}
\caption{Dependence of $r^2|\Psi(r)|^2$ on $r$ for
the three coupling regimes. Color-coded $\alpha$ values: 
0.0 (black), 0.1 (red), 0.2 (blue), 0.3 (magenta).}
\end{minipage}\hspace{2pc}%
\begin{minipage}{18pc}
\includegraphics[width=17.7pc,height=16.8pc]{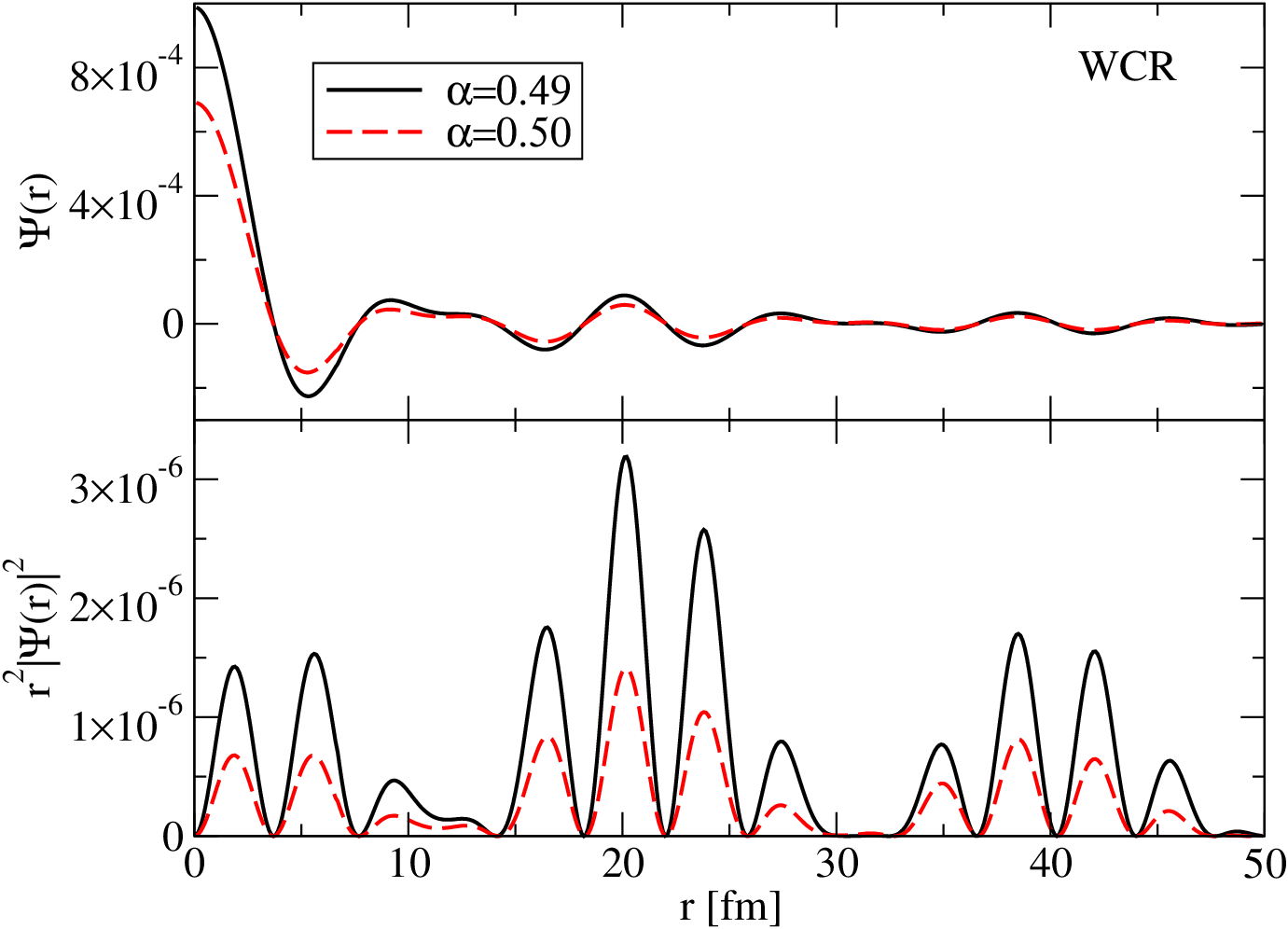}
\caption{Dependence of $\Psi(r)$ and $r^2|\Psi(r)|^2$
on $r$ in the WCR for two asymmetries at which the
LOFF phase is the ground state.  }
\end{minipage} 
\end{figure}

\subsection{Occupancies, Excitations, and Pauli Blocking Effects}

Analysis of the kernel $K(k)$, the $p$ and $n$ occupation probabilities,
and quasiparticle dispersion relations reveal prominent effects of 
Pauli blocking (the ``breach''). Summarizing the phenomena 
occurring in the different coupling regimes, we find

\begin{itemize}
\item[$\bullet$]
WCR: At large asymmetries, the minority component is expelled 
from the blocking region ($n_p \approx 0$), while the majority 
component is maximally occupied ($n_n/2 \approx 1$).  The ``breach''
is filled in with increasing $T$.  

\item[$\bullet$]
WCR: The LOFF phase appearing in this regime largely mitigates
the blocking mechanism by allowing for a non-zero CM momentum
of the condensate.  Accordingly, all intrinsic properties are 
much closer to those of the isospin-symmetric BCS phase.

\item[$\bullet$]
WCR: In the small-$\alpha$ limit, the occupation numbers are clearly 
fermionic, with some diffuseness due to the temperature.

\item[$\bullet$]
ICR: The fermionic nature of the occupation numbers is lost, 
a Fermi surface cannot be identified, and no ``breach'' appears.

\item[$\bullet$]
ICR: For large $\alpha$ values, the occupation numbers become 
non-monotonic; for the minority component this is a precursor 
of a change in the topology of the Fermi surface in the transition 
ICR $\rightarrow$ SCR.

\item[$\bullet$]
SCR: the occupation numbers and other properties are consistent
with a BEC of strongly-coupled pairs (deuterons).  At large asymmetries, 
the Fermi sphere of the minority component in the WCR has evolved into 
a shallow shell structure.

\item[$\bullet$]
SCR: long-range coherence of the condensate is lost.

\item[$\bullet$]
WCR $\rightarrow$ ICR $\rightarrow$ SCR: The quasiparticle dispersion 
relation changes in form from that corresponding to the existence
of a Fermi surface to one that is minimal at $k=0$, independent
of isospin asymmetry.

\item[$\bullet$]
With increasing $\alpha$, the proton component acquires 
points with zero excitation energy, as in gapless superconductivity.
The occupation numbers reach a maximum at finite $k$ 
and reflect a change of topology: from the filled Fermi sphere
to one with an empty ``core.''
\end{itemize}

\section{The Missing Ingredient: Formation of Nuclear Clusters}

The matter in supernova envelopes is (i) at finite isospin asymmetry
(though remaining quite small compared to its values in the crust 
of a neutron star), and (ii) at relatively high temperatures compared 
to those considered in the present study.  This environment
could support the existence of a substantial population of nuclear 
clusters besides deuterons, notably tritons, $^3$He nuclei, and 
alpha particles.  The $\alpha$ particles may form a BEC at 
sufficiently low temperatures.  The extent to which the presence 
of such clusters will modify the structure of the phase diagram 
constructed so far remains to be determined at a comparable
level of microscopic precision.  Additionally, formation of 
neutron dimers in the nuclear medium remains a tantalizing 
possibility, which will be explored schematically in the
second part of this review.

\section{ Framework for BCS Pairing Versus Hidden Dimer State: ``Take One''}

The $NN$ interaction in the $^1S_0$ partial wave, having a scattering
length $a = -18.95~{\rm fm}$ and a strong resonance at the tiny energy
of 0.067 MeV, just barely fails to support a bound state.  Although
this hypothetical dineutron cannot exist in vacuum, there remains
a tantalizing potential for it to be bound in a nuclear medium
and even in pure neutron matter.  For example, it would take an effective,
in-medium mass $M^*$ of about 5\% over the bare mass $M$ for a dineutron
to bind in the pure neutron system.  One could think of such a
dineutron as a nuclear analog of the polaron of solid-state physics,
which also has no existence in vacuum.

Recently, Eckhard Krotscheck's quantum many-body group at Buffalo
\cite{fankro} has developed and executed a powerful {\it ab initio}
approach within correlated-basis theory that is capable of quantitative
prediction of the ground-state properties of strongly interacting
Fermi systems in the low-density regime, with errors in the
ground-state energy below 2\%.  For two-body interactions qualitatively
similar to the $S$-wave neutron-neutron potential, these authors have
exploited rigorous variational aspects of this approach to demonstrate
the existence of a robust singularity at low (but not asymptotically
low) density that is plausibly associated dimerization of the system
driven by phonon exchange.  This behavior is analogous to what occurs
at zero density when the in-vacuum scattering length diverges;
accordingly, it is interpreted in terms of divergence of a well-defined
in-medium scattering length at finite density.  On a more phenomenological
plane, it is important to highlight not only historical \cite{feather}
early interest in possible existence of a dineutron, but more
significantly an extensive body of recent theoretical and experimental
work \cite{dn1-8}, respectively predicting and providing evidence
for strong dineutron correlations in finite nuclei or nuclear matter,
even suggesting a tendency toward Bose-Einstein condensation of
long-lived dineutrons.

Our objective for this section of the review and the next is to 
furnish a schematic framework \cite{BE,vak} for the emergence of 
a hidden dimer state in homogeneous Fermi matter (potentially a dineutron 
in the case of realistic nuclear matter), which competes energetically 
with BCS pairing states, both conventional and unconventional.  The 
present analysis follows the same paths as explored in \cite{BE}, 
but uncovers additional subtleties beyond what is revealed in that
work.  In some important aspects, this treatment supplements 
the {\it ab initio} study carried out in \cite{fankro} within the 
method of correlated basis functions.  The analysis is primarily 
based on the Thouless criterion for determination of the critical 
temperature for the onset or termination of pairing correlations.  The 
Thouless criterion \cite{thouless,urban} centers on the behavior of the 
linear integral equation
\begin{equation}
{\cal T}(p,T)=-\int V(p,p_1){\tanh\left(\epsilon(p_1)/2T\right)
\over 2\epsilon(p_1)}{\cal T}(p_1,T)d\upsilon_1 .
\label{thc}
\end{equation}
for the finite-temperature $T$-matrix (denoted here by $\cal T$),
stating that if it does not have a pole at temperature $T$, this
temperature lies above above the transition temperature.  The
single-particle spectrum appearing in this equation is given by
$\epsilon(p) = p^2/2M - \mu$, with $M$ the bare mass and $\mu$
the chemical potential.

Analysis of Eq.~(\ref{thc}) is expedited by decomposing the momentum-space
pairing interaction $V(p_1,p_2)$ identically into a separable component
and a remainder $R(p_1,p_2)$ that vanishes when either momentum argument
is on the Fermi surface:
\begin{equation}
V(p_1,p_2)= V_F\phi(p_1)\phi(p_2)+ R(p_1,p_2) .
\label{decomp}
\end{equation}
The latter property is guaranteed by the choice $\phi(p)= V(p,p_F)/ V_F$,
with $ V_F= V(p_F,p_F)$.  Inserting this expression into Eq.~(\ref{thc}),
simple algebra leads to an equivalent set of two coupled equations.
The first is a linear integral equation for the shape 
$\chi(p) = {\cal T}(p)/{\cal T}(p_F)$ of ${\cal T}$ in momentum space:
\begin{equation}
\chi(p)=\phi(p)-\int R( p,p_1) { \tanh\left(\epsilon(p_1)/2T\right)
\over 2\epsilon(p_1)}\chi(p_1)d\upsilon_1,
\label{first}
\end{equation}
while the second equation takes the form
\begin{equation}
-{1\over V_F}=\int \phi(p){\tanh\left(\epsilon(p)/2T\right)
\over 2\epsilon(p)}\chi(p)d\upsilon.
\label{second}
\end{equation}
Due to the imposed behavior of the remainder function ${\cal R}$,
the description of singular behavior in the Cooper pairing channel
is isolated in the second equation.  The same tactic was invoked to 
develop an efficient and accurate procedure \cite{kkc} for solving 
BCS-type gap equations in those cases where the pairing interaction 
contains a strong inner repulsion in coordinate space.  Widely used 
versions of the $NN$ interaction, such as the Reid soft-core and 
Argonne V18, are of this character.

Introducing the difference $\eta(p)=\chi(p)-\phi(p)$, the second
equation may be rewritten as
\begin{equation}
-{1\over V_F}= I_{11}(T) +\int \phi(p) {\tanh\left(\epsilon(p)/2T\right)
\over 2\epsilon(p)}\eta(p)d\upsilon,
\label{third}
\end{equation}
where
\begin{equation}
I_{11}(T)=\int \phi(p){\tanh \left(\epsilon(p)/2T\right)
\over 2\epsilon(p)}\phi(p)d\upsilon,
\end{equation}
while $\eta(p)$ obeys an integral equation replacing (\ref{first}), viz.\
\begin{equation}
\eta(p,T)=
-\int R( p,p_1) {\tanh \left(\epsilon(p_1)/2T \right) \over
2\epsilon(p_1)}\left(\phi(p_1)+\eta(p_1,T)\right)\,d\upsilon_1 .
\label{etaeq}
\end{equation}
At the bifurcation point $T=0$, the first term on the right side
of Eq.~(\ref{third}), i.e., $I_{11}(T)$, diverges logarithmically,
behaving essentially as $0.5N(0)\ln (\epsilon_c/T)$.
Therefore a solution $T=0$ exists only if the second term
{\it also} diverges at this point.

To confirm that a compensating divergence occurs, we expand
the function $\eta(p)$ in a basis formed by the eigenfunctions
$\zeta_n(p)$ of the kernel $R(p,p';\rho)L(p,0)$ involving
the propagator
\begin{equation}
L(p,\omega) = -(1-2n(p))/(\omega -2\epsilon(p)-i\delta\,{\rm sgn}(p-p_F)).
\label{ldef}
\end{equation}
The expansion is dominated by the
contribution from the eigenfunction $\zeta_0(p)$ belonging to the
lowest eigenvalue $\sigma_0$ of this kernel, which satisfies
\begin{equation}
\zeta_0(p)=-\sigma_0\int R(p,p_1,\rho){1\over 2|\epsilon(p_1)|}
\zeta_0(p_1) d\upsilon_1.
\label{eigenvaleq}
\end{equation}
Thus we write
\begin{equation}
\eta(p,T) =  \eta_0(T)\zeta_0(p) + \vartheta(p),
\label{vartheta}
\end{equation}
where the remainder $\vartheta(p)$ vanishes on the Fermi surface.
Inserting this expression into the integral equation (\ref{etaeq})
for $\eta(p,T)$ and collecting all terms having $\eta_0(p,T)$ as a
factor on the left, we have
\begin{equation}
\eta_0(T)\left(\zeta_0(p) +
\int\, R(p,p_1){{\tan(\epsilon(p_1)/2T)}\over{2\epsilon(p_1)}}
\zeta_0(p_1)\,d\upsilon_1\right) = Z(p),
\label{etaZ}
\end{equation}
where
\begin{equation}
Z(p) = -\vartheta(p) -
\int\,  R(p,p_1){{\tan(\epsilon(p_1)/2T)}\over{2\epsilon(p_1)}}
(\phi(p_1) + \vartheta(p_1))\,d\upsilon_1.
\label{Zeq}
\end{equation}

It is convenient to use the eigenvalue equation (\ref{eigenvaleq})
for $\zeta_0$ to rewrite the left side of Eq.~(\ref{etaZ}) as
\begin{equation}
\eta_0(T)\left ({\kappa\over \sigma_0}\zeta_0(p)+
\int\,  R(p,p_1){\cal D}(p_1,T) \zeta_0(p_1)d\upsilon_1\right)=Z(p) ,
\end{equation}
where
\begin{equation}
{\cal D}(p,T)={ \tanh\left(\epsilon(p)/2T\right) \over
2\epsilon(p)}-{1\over 2|\epsilon(p)|}
\end{equation}
and $\kappa \equiv \sigma_0 -1$.
Next, both sides of this result are multiplied by the product
$\zeta_0(p)/(2|\epsilon(p)|)$, and the momentum integration is
performed.  The operator $R$ may be eliminated from (\ref{etaZ}) with
the aid of (\ref{eigenvaleq}) to obtain
\begin{equation}
\eta_0(T)=(\kappa+\gamma(T))^{-1}I_{10}/I_{00},
\label{main}
\end{equation}
where $I_{00}>0$ is given by
\begin{equation}
I_{00} = \int\, \zeta_0(p)L(p,0)\zeta_0(p)\,d\upsilon,
\label{I00}
\end{equation}
while
\begin{equation}
\gamma(T)=I_{00}^{-1}\int\,\zeta_0(p) \left({\tanh (\epsilon(p)/2T)
\over 2\epsilon(p)}-{1\over 2|\epsilon(p)|}\right)\zeta_0(p)\,d\upsilon  
\label{gammaeq}
\end{equation}
and
\begin{equation}
I_{10}=\sigma_0\int \zeta_0(p){1\over 2|\epsilon(p)|} Z(p) \, d\upsilon .
\end{equation}
Equation (\ref{main}) is the main result of this analysis.

To demonstrate that the sign of $\gamma(T)$ is positive, as required,
the integrand in Eq.~(\ref{gammaeq}) is rearranged to write
\begin{equation}
\gamma(T)\propto \int\zeta_0^2(p){\tanh ({|\epsilon(p)|/ 2T})
-1\over 2|\epsilon(p)|} d\upsilon .
\end{equation}
Since $\zeta_0(p)$ vanishes at the Fermi surface like $\epsilon(p)$,
we may conclude that $\gamma(T)= \gamma T^2$, where henceforth
$\gamma$ is a positive constant.

Considering next the integral $I_{10}$, the explicit form for
$Z(p)$ is inserted into its integrand.  Noting that that the terms
involving the remainder $\vartheta$ essentially cancel one another,
we may then take
\begin{equation}
I_{10}=\int \zeta_0(p) {1\over 2|\epsilon(p)|}\phi(p)d\upsilon .
\end{equation}
Since $\gamma(T)$ vanishes as $T^2$ for $T\to 0$, we see from
Eq.~(\ref{main}) that the coefficient $\eta_0(T)$ does in fact
diverge together with the function $\eta(p)$ itself at the
critical point where $\kappa = \sigma_0 -1$ changes sign.

We are now ready to substitute the result (\ref{main}) into the
dispersion relation (\ref{third}).
Implementing an energetic cutoff $\epsilon_c \sim \epsilon_F^0$
in the result, we then arrive at
\begin{equation}
{1 \over 2} \ln (\epsilon_c/T)={1 \over \lambda}
-{\nu^2\over \kappa(\rho)+\gamma T^2},
\label{skel}
\end{equation}
where it is to be understood that all minor corrections are included in the
effective pairing constant denoted $\lambda$.

This analysis was performed with the case $(\lambda > 0,\,\kappa < 0)$
in mind.  Near the critical density $\rho_t$, the ``stiffness''
parameter $\kappa$, being small and negative, behaves as
\begin{equation}
\kappa \sim -| (\partial \kappa /\partial \rho)_t |(\rho_t -\rho).
\end{equation}
For small  $|\rho_t-\rho|$, the contribution of the $1/\lambda$ term on 
the right side of Eq.~(\ref{skel}) can be neglected, and we find
\begin{equation}
     T_1 \propto \exp[-(a \rho_t)/(\rho_t-\rho)],
\end{equation}
where $a$ is a numerical constant.  Since the BCS constant $\lambda$ does 
not enter this result, we infer that the solution so obtained does not 
belong to the BCS type.

However, there is in fact a second root of Eq.~(\ref{skel}), lying
at temperatures $T_2>\sqrt{|\kappa|/\gamma}>T_1$.  To assess its 
location, we set $\kappa=0$ in Eq.~(\ref{skel}) to 
write
\begin{equation}
{\lambda\gamma \over 2} T^2 \ln (\epsilon_c/T)= \gamma T^2-\lambda\nu^2.
\end{equation}
The solution of this equation has the form
\begin{equation}
{T_2}^2=\lambda\nu^2/\gamma+ O(\lambda^2\ln\lambda).
\label{skel1}
\end{equation}
Since the result (\ref{skel1}) depends explicitly on the BCS constant 
$\lambda$, it is this solution that corresponds to Cooper pairing.  We 
conclude that the dineutron state loses in its competition with the BCS-like 
state (\ref{skel1}), as long as the magnitude of the critical parameter 
$|\kappa|$ remains small.

The situation at $T\to 0$ changes completely when $|\kappa|$ increases
beyond $\lambda\nu^2$, for then the first term on the right side of
Eq.~(\ref{skel}) prevails over the second.  As a result, the first root 
of this equation has the standard BCS form $T_1=\epsilon_c \exp[-1/\lambda]$. 
However, as before there exists a different and larger root, determined 
by the relation
\begin{equation}
T_2=T_{nn}= \sqrt{(|\kappa|+\lambda\nu^2)/\gamma}\simeq \sqrt{|\kappa|/\gamma},
\end{equation}
implying that the corresponding solution is of dineutron character. Thus, 
at sufficiently large $|\kappa|$ the dineutron state wins the contest
for survival.

\begin{figure}[tb]
\begin{center}
\includegraphics[width=12truecm]{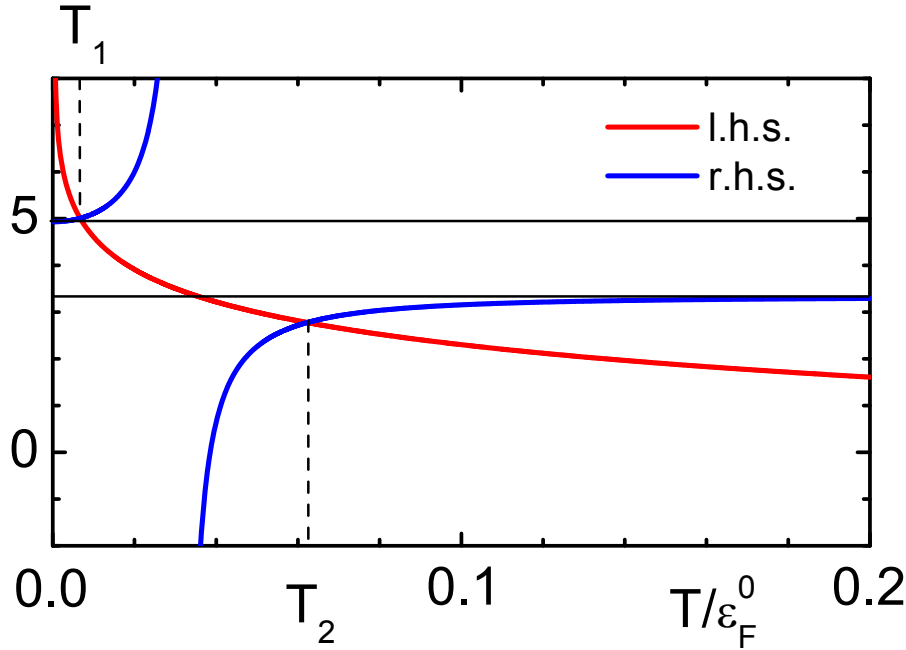}
\caption{
Graphical demonstration of two solutions of Eq.~(\ref{skel}).
Model parameters adopted: $\lambda = 0.3$, $\nu = 0.04$, $\kappa = - 0.001$,
$\gamma = (\epsilon_F^0)^{-1/2}$, $\epsilon_c =\epsilon_F^0$.
The in-medium dineutron solution rather than the BCS solution ensures
the maximum critical temperature for termination of pairing correlations
provided $|\kappa|>\lambda\nu^2$; otherwise the BCS solution still holds.
}
\label{fig:DiNeutron1}
\end{center}
\end{figure}

Fig.~\ref{fig:DiNeutron1} shows a graphical solution of the key relation
(\ref{skel}) derived above, the running variable $T$ being measured in units
of $\epsilon_F^0=p_F^2/2M$.  The left side of this equation, plotted
versus $T$, is traced by the red line, while the right side is drawn as two
disjoint blue lines. Model inputs are stated in the figure caption. The two
crossing points determine two critical temperatures, $T_1$ and $T_2$,
with $T_2 > T_1 $.

\section{Framework for BCS Pairing Versus Hidden Dimer State: ``Take Two''}

We now expand the analysis from the case $(\lambda > 0,\,\kappa < 0)$
to the other quadrants of the $\lambda-\kappa$ parameter plane.
The strategy now is to track the location of the Cooper singlet-channel
pole of the zero-temperature scattering amplitude in the normal state
of the homogeneous many-fermion system.  We pursue this task based on
the Bethe-Salpeter equation for the corresponding vertex part
${\cal K}_{\alpha \beta}({\bf p},\omega)
= {\cal K}({\bf p},\omega)(\tau_2)_{\alpha \beta}$.  Here $\alpha$,
$\beta$ are spin indices, ${\bf p}$ is the momentum of the incoming
particle (its target having momentum $-{\bf p}$), and $\omega$
is the total two-particle energy measured from $2 \mu$.  The
equation ready for analysis now reads
\begin{equation}
{\cal K}(p,\omega) = -\int V(p,p_1) L(p_1,\omega)
{\cal K}(p_1,\omega)\, d\upsilon_1,
\label{bethesalt}
\end{equation}
where $V(p,p_1)$ represents the block of diagrams irreducible in the
particle-particle channel (here just the pairing interaction) and
$L(p,\omega)$, again given by Eq.~(\ref{ldef}), is the particle-particle
propagator in the normal ground state. The analysis of the integral
equation (\ref{bethesalt}) now proceeds essentially in parallel with 
that carried out for Eq.~(\ref{thc}). We bypass the lengthy details and 
move immediately to the final result for the basic dispersion equation, 
whose analytic continuation to the complex-$\omega$ plane yields
\begin{equation}
{1\over 2} \left( \ln {\epsilon_c \over \omega} + i {\pi \over 2}\
\right) = {1 \over \lambda} - {\nu^2 \over {\kappa + B(\omega)}},
\label{disp}
\end{equation}
where $\nu^2 = I_{10}^2 /I_{00} N(0)$, with $N(0)$ denoting the
normal density of states.  The function $B(\omega)$ is given by
\begin{equation}
B(\omega) = - (I_{00})^{-1} \int \zeta_0(p) \delta L(p,\omega)\zeta_0(p) \,
d \upsilon,
\end{equation}
with $\delta L(p,\omega) = L(p,\omega) - L(p,0)$.
Setting $\Omega = i \omega$ in Eq.~(\ref{disp}), we arrive at the
general formula for determination of the pairing-gap proxy $\Omega$,
\begin{equation}
{1 \over 2} \ln {\epsilon_c \over \Omega} = {1 \over \lambda}
- {\nu^2 \over{\kappa + B \Omega^2 \ln(\epsilon_c/\Omega)}}
\label{general}
\end{equation}
with $B>0$.

These results prepare us to uncover systematically the nature of pairing
solutions in relevant domains of the $\lambda-\kappa$ plane, while
acknowledging that in a {\it realistic} two-nucleon interaction these
two parameters actually constrain one another. (We note that the
quadrant $(\lambda < 0,\, \kappa > 0)$ is empty of solutions since
the right side of Eq.~(\ref{general}) is never positive.)

\begin{itemize}
\item[$\bullet$]
{\it Quadrant} $(\lambda > 0,\, \kappa > 0)$.
The effect of the second term on the right side of Eq.~(\ref{general}), 
proportional to $\nu^2$, amounts just to a renormalization of the 
coupling constant $\lambda$ in the first term.  Only the BCS solution 
survives in this case.

\item[$\bullet$]
{\it Quadrant} $(\lambda > 0,\, \kappa < 0)$.  This case has been dealt 
with earlier in the context of the Thouless criterion.  The parameter 
$\kappa$ goes negative at densities below some threshold density $\rho_t$, 
thereby triggering the onset of the dineutron state.  As already 
established above, two different solutions exist in this quadrant, 
a BCS solution and another corresponding to a hidden dineutron state.  
It is informative to track the trajectories of both roots $\Omega_{1,2}$ 
as functions of $\lambda$.  As $\lambda$ goes to zero, the left root 
(nearest the origin) has the behavior $\Omega_1 \propto e^{-2/\lambda}$,
with $\Omega_1(0) = 0$.  This is obviously the ordinary BCS root.
In the same limit, the other root, $\Omega_2$, is situated close
to $\sqrt{{|\kappa|}/B}$ and is clearly of non-BCS character.  With
increasing $\lambda$, both roots move away from the origin.  It should
be emphasized that in the analysis performed, both parameters $\lambda$
and $\kappa$ are assumed to remain small to assure the smallness of
the roots, but with this proviso the analysis performed is self-consistent.
As the roots evolve with increasing $\lambda$, the two solutions switch 
ascendancy in the region where $\lambda\nu^2\simeq |\kappa|$. Thus we 
find that the dineutron state wins the competition with the BCS state
only if the BCS coupling constant is rather small.

\item[$\bullet$]
{\it Quadrant} $(\lambda < 0,\, \kappa < 0)$.
Here only the root
\begin{equation}
T^2_{nn}=(|\kappa|-|\lambda|\nu^2)/\gamma
\end{equation}
survives, but it exists only in the limit of sufficiently large  
$|\kappa|>|\lambda|\nu^2$.  Evidently, this solution is of dineutron 
character.

\end{itemize}

\section{Conclusions and Outlook}

In this mini-review we have discussed some recent (and not so recent)
advances toward the goal of a comprehensive understanding of the phase
diagram of nuclear matter in the low-density regime below about half
the saturation density of heavy nuclei.  In this density region one
may safely regard neutrons and protons as the basic constituents,
interacting through two-nucleon potentials constrained by
nucleon-nucleon scattering data and binding energies of light nuclei
(deuteron, triton, etc.).  An {\it ab initio} many-body approach has
been applied in a quantitative exploration of the temperature-density
($T-\rho$) phase diagram of this apparently well-defined quantum
many-body system \cite{II}, focusing (a) on its dependence upon isospin
asymmetry and (b) simultaneously on the roles played by pairing and the
BCS-BEC crossover from Cooper pairs to bound bosonic dimers
(deuterons). In addition to conventional BCS pairing, its competition
with unconventional LOFF pairing (in which the Cooper pairs have
finite total momentum) has been treated in some detail.  Away from
isospin symmetry, the heterogeneous phase which combines symmetrical
superfluid and asymmetric normal matter may have lowest free energy.
This possibility has been taken into consideration in the numerical
calculations, with the finding that two-phase mixtures are favored
over the homogeneous phases at the lowest temperatures and
no-to-high density, where one of the components accommodates
the excess neutrons.

The major portion of this review (Secs.~2-10) has dealt with
the $T-\rho$ phase diagram of dilute nuclear matter in its
dependence on isospin.  In Secs.~12 and 13 the focus has been
shifted to another intriguing aspect of the nuclear-matter problem.
Restricting attention to pure neutron matter for simplicity,
this is the possibility of unconventional pairing involving the
emergence of neutron dimers, i.e., dineutrons, in some density
range (obviously not including $\rho \to 0$). In turn, this eventuality
raises the question of whether there is a corresponding BCS-BEC
transition in pure neutron matter.

A very recent study carried out by some of us \cite{III} has a bearing
on these issues. Our initial study concentrated on the $T$-$\rho$
phase diagram of neutron matter including only homogeneous and
isotropic phases indexed by a spin-polarization parameter. The
resulting diagram resembles that of asymmetrical nuclear matter
because of the analogy between nuclear matter at non-zero isospin
asymmetry and pure neutron matter at non-zero spin asymmetry.  An
essential distinction between the phase diagrams of these systems
stems from the fact that two neutrons are not bound in vacuum,
whereas a neutron and a proton bind into the deuteron in free
space.  Accordingly, pure neutron matter cannot, in the strict sense,
undergo a BCS-BEC crossover from the $^1S_0$ BCS superfluid existing
at higher densities, to a BEC condensate of neutron dimers at
asymptotically low density.  The end state must be a neutron
gas. Nevertheless, the many-body calculations carried out recently
by Stein et al.~\cite{III} give clear evidence of a BCS-BEC {\it
precursor} and thus confirm earlier findings of various authors
obtained at zero spin polarization (see \cite{III} for references).
The existence of a BCS-BEC precursor is revealed, for example, by
the attainment of Cooper-pair correlation lengths comparable to
the interparticle distance.  This behavior must be attributed to
the strongly resonant character of the $nn$ interaction at low
densities, where the $^1S_0$ component is dominant.

This discourse impels us to ponder the content of Secs.~12
and 13, which provide a schematic framework for understanding
the emergence of unconventional types of pairing (notably, a
``hidden dineutron state'') along with the conventional BCS solution.
A systematic classification of such solutions is achieved in
terms of two parameters: a familiar coupling parameter $\lambda$
and a so-called stiffness parameter $\kappa$.  As introduced,
these parameters would be responsible for subsuming the effects
of such complications relative to basic, bare-interaction nuclear
BCS theory as (i) medium-induced corrections to the in-vacuum pairing
interaction (coming from exchange of density and spin-density
fluctuations, for example) and (ii) non-trivial momentum
dependence of the resulting dressed interaction.  (Attention
should also be given to self-energy corrections.) The first category
of effects (``polarization corrections'') is absent in the many-body
treatment underlying the results presented in Secs.~2-10 and in
\cite{II,III}.

The recent, seminal work of Fan et al.~\cite{fankro} does consistently
incorporate such effects in a quantitatively precise {\it ab initio}
many-body approach developed within correlated basis theory.  This
approach has been exhaustively applied in a study correlations in
the low-density, strongly interacting Fermi gas that addresses
its Fermi-liquid state and BCS pairing, as well as dimerization
induced by phonon exchange.  Application of this method to neutron
matter has the potential of resolving the fate of the dineutron,
which may well depend (perhaps counterintuitively) on the hardness
of the inner repulsion of the ${^1}S_0$ neutron-neutron interaction.

\section*{Acknowledgments}
MS acknowledges support from the HGS-HIRe graduate program at Frankfurt 
University. AS is supported by the Deutsche Forschungsgemeinschaft 
(Grant No.~SE 1836/3-1) and by the NewCompStar COST Action MP1304. 
XGH is supported by Fudan University Grant EZH1512519 and Shanghai 
Natural Science Foundation Grant No.14ZR1403000. Additionally, this 
research was partially supported by RFBR grants 13-02-00085, 14-02-00107, 
and 15-02-06261, and by grant NS-932.2014.2 from the Russian Ministry 
of Sciences.  VAK thanks the McDonnell Center for the Space Sciences 
for timely support.  JWC is indebted to the University of Madeira and 
its Centro de Ci\^encias Matem\'aticas for gracious hospitality 
during his sabbatical residency.

\end{document}